\documentclass{emulateapj}
\usepackage{apjfonts}
\newcommand{\pivec}{\mbox{\boldmath $\pi$}}

\lefthead{JUNG ET AL.}
\righthead{ }

\begin{document}
\title{OGLE-2016-BLG-1003: First Resolved Caustic-crossing Binary-source Event 
Discovered by Second-generation Microlensing Surveys  
}

\author{
Y.~K.~Jung$^{1,22}$, 
A.~Udalski$^{2,23}$, 
I.~A.~Bond$^{3,24}$,
J.~C.~Yee$^{1,22}$ \\
and \\
A.~Gould$^{4,5,6}$, C.~Han$^{7}$, M.~D.~Albrow$^{8}$, C.-U.~Lee$^{4,9}$, S.-L.~Kim$^{4,9}$, 
K.-H.~Hwang$^{4}$, S.-J.~Chung$^{4,9}$, Y.-H.~Ryu$^{4}$, I.-G.~Shin$^{1}$, W.~Zhu$^{5}$, 
S.-M.~Cha$^{4,10}$, D.-J.~Kim$^{4}$, Y.~Lee$^{4,10}$, B.-G.~Park$^{4,9}$, H.-W.~Kim$^{4}$, 
R.~W.~Pogge$^{5}$ \\ 
(The KMTNet Collaboration) \\
J.~Skowron$^{2}$, M.~K.~Szyma{\'n}ski$^{2}$, R.~Poleski$^{2,5}$, P.~Mr{\'o}z$^{2}$, 
S.~Koz{\l}owski$^{2}$, P.~Pietrukowicz$^{2}$, I.~Soszy{\'n}ski$^{2}$, K.~Ulaczyk$^{2}$, 
M.~Pawlak$^{2}$ \\
(The OGLE Collaboration) \\
F.~Abe$^{11}$, D.~P.~Bennett$^{12,13}$, R.~Barry$^{14}$, T.~Sumi$^{15}$, 
Y.~Asakura$^{11}$, A.~Bhattacharya$^{12}$, M.~Donachie$^{16}$, A.~Fukui$^{17}$, 
Y.~Hirao$^{15}$, Y.~Itow$^{11}$, N.~Koshimoto$^{15}$, M.~C.~A.~Li$^{16}$, 
C.~H.~Ling$^{18}$, K.~Masuda$^{11}$, Y.~Matsubara$^{11}$, Y.~Muraki$^{11}$, 
M.~Nagakane$^{15}$, N.~J.~Rattenbury$^{16}$, P.~Evans$^{16}$, A.~Sharan$^{16}$, 
D.J.~Sullivan$^{19}$, D.~Suzuki$^{13}$, P.~J.~Tristram$^{20}$, T.~Yamada$^{15}$, 
T.~Yamada$^{21}$, A.~Yonehara$^{21}$ \\ 
(The MOA Collaboration)
}

\bigskip\bigskip
\affil{$^{1}$Smithsonian Astrophysical Observatory, 60 Garden St., Cambridge, MA 02138, USA}
\affil{$^{2}$Warsaw University Observatory, Al. Ujazdowskie 4, 00-478 Warszawa, Poland}
\affil{$^{3}$Institute of Natural and Mathematical Sciences, Massey University, Auckland 0745, New Zealand}
\affil{$^{4}$Korea Astronomy and Space Science Institute, Daejon 305-348, Republic of Korea}
\affil{$^{5}$Department of Astronomy, Ohio State University, 140 W. 18th Ave., Columbus, OH 43210, USA}
\affil{$^{6}$Max-Planck-Institute for Astronomy, K$\rm \ddot{o}$nigstuhl 17, 69117 Heidelberg, Germany}
\affil{$^{7}$Department of Physics, Chungbuk National University, Cheongju 371-763, Republic of Korea}
\affil{$^{8}$University of Canterbury, Department of Physics and Astronomy, Private Bag 4800, Christchurch 8020, New Zealand}
\affil{$^{9}$Korea University of Science and Technology, 217 Gajeong-ro, Yuseong-gu, Daejeon 34113, Korea}
\affil{$^{10}$School of Space Research, Kyung Hee University, Yongin 446-701, Republic of Korea}
\affil{$^{11}$Institute for Space-Earth Environmental Research, Nagoya University, Nagoya 464-8601, Japan}
\affil{$^{12}$Department of Physics, University of Notre Dame, Notre Dame, IN 46556, USA}
\affil{$^{13}$Laboratory for Exoplanets and Stellar Astrophysics, NASA/Goddard Space Flight Center, Greenbelt, MD 20771, USA}
\affil{$^{14}$Astrophysics Science Division, NASA Goddard Space Flight Center, Greenbelt, MD 20771, USA}
\affil{$^{15}$Department of Earth and Space Science, Graduate School of Science, Osaka University, Toyonaka, Osaka 560-0043, Japan}
\affil{$^{16}$Department of Physics, University of Auckland, Private Bag 92019, Auckland, New Zealand}
\affil{$^{17}$Okayama Astrophysical Observatory, National Astronomical Observatory of Japan, 3037-5 Honjo, Kamogata, Asakuchi, Okayama 719-0232, Japan}
\affil{$^{18}$Institute of Information and Mathematical Sciences, Massey University, Private Bag 102-904, North Shore Mail Centre, Auckland, New Zealand}
\affil{$^{19}$School of Chemical and Physical Sciences, Victoria University, Wellington, New Zealand}
\affil{$^{20}$Mt. John University Observatory, P.O. Box 56, Lake Tekapo 8770, New Zealand}
\affil{$^{21}$Department of Physics, Faculty of Science, Kyoto Sangyo University, 603-8555 Kyoto, Japan}
\footnotetext[22]{The KMTNet Collaboration.}
\footnotetext[23]{The OGLE Collaboration.}
\footnotetext[24]{The MOA Collaboration.}

\begin{abstract}
We report the analysis of the first resolved caustic-crossing binary-source 
microlensing event OGLE-2016-BLG-1003. The event is densely covered by the 
round-the-clock observations of three surveys. The light curve is characterized 
by two nested caustic-crossing features, which is unusual for typical caustic-crossing 
perturbations. From the modeling of the light curve, we find that the anomaly 
is produced by a binary source passing over a caustic formed by a binary lens. 
The result proves the importance of high-cadence and continuous observations, 
and the capability of second-generation microlensing experiments 
to identify such complex perturbations that are previously unknown. 
However, the result also raises the issues of 
the limitations of current analysis techniques for understanding lens systems 
beyond two masses and of determining the appropriate multiband observing 
strategy of survey experiments. 
\end{abstract}

\section{Introduction}

Since the first microlensing surveys \citep{alcock93,aubourg93,udalski93}, 
the microlensing experiment has achieved remarkable progress through the advent of 
new observation surveys (e.g., MOA: \citealt{sumi11}, KMTNet: \citealt{kim16}), 
and the improvement in both software (e.g., improved photometry based on difference 
imaging) and hardware (e.g., large-format wide-field cameras). As a result, 
the detection rate of microlensing events, which was several dozen per year 
in early experiments, is now more about two thousand events per year. In addition, 
the progress now enables surveys to densely and continuously cover lensing events 
with a high enough cadence to detect various microlensing signals without 
follow-up observations.

Dense and continuous coverage of lensing light curves is important for various 
scientific studies in microlensing. The most prominent example is planetary 
microlensing events. The lensing signal of a planet generally appears
a short-lasting perturbation to a standard \citet{paczynski86} curve induced 
by the host of the planet, and the duration of the signal is on the order of 
hours for Earth-mass planets and several days for Jupiter-mass planets. 
Therefore, one would expect that lensing events need to be monitored with 
a cadence of $\lesssim 1~\rm hr^{-1}$ for low-mass planets and of 
$\lesssim 1~\rm day^{-1}$ for giant planets. With the improvement of observational 
equipment and the development of observational strategy, current-generation 
experiments are able to monitor large number of lensing events with cadences high 
enough to detect planets by survey only data (e.g., \citealt{udalski15c,sumi16,shin16}).

Another area that has benefitted from dense and continuous observations is 
caustic-crossing microlensing events. When a lens is composed of multiple masses, 
the light curve often exhibits a slope discontinuity. 
This discontinuity occurs when a source is located 
at some positions (caustics) at which the point-source magnification diverges 
\citep{schneider86,erdl93}. \footnote{It is possible that observed light curves 
are continuous due to smearing out of the magnification by the finite size of the source, 
but the derivative is discontinuous (Figure 1 from \citealt{gould99}).} 
In addition, if the source passes close to or over the caustic, 
the different parts of the source are magnified differently, resulting in the deviation 
from the point-source magnification. Detecting this finite-source effect is important 
because it provides an opportunity to measure the angular Einstein radius $\theta_{\rm E}$, 
which makes it possible to better constrain the physical properties of the lens system 
\citep{gould94}. Despite its importance, the measurement of $\theta_{\rm E}$ is difficult 
due to the short duration of caustic crossings as well as the difficulty of predicting 
the time of occurrence. For most caustic-crossing events, the duration of the signal 
is order of hours. Therefore, resolving the caustic crossings also requires high-cadence 
observations similar to the cadence for low-mass planets \citep{shin16}.

High-cadence and continuity of microlensing observations is also crucial to identify 
the nature of lensing events. Aside from inherent degeneracies originating in symmetries 
of the lens-mapping equation itself (e.g., \citealt{dominik99}), one often confronts 
cases in which different interpretations can explain 
observed light curves due to the incomplete coverage of the perturbation. For example, 
\citet{gaudi04} showed that sparse coverage of planet-like anomalies can give rise to 
severe ambiguities in the interpretation. In addition, \citet{park14} pointed out that 
incomplete coverage also causes ambiguity between stellar and planetary 
interpretations even though the perturbation clearly shows a large deviation 
from the standard Paczy\'{n}ski curve. Furthermore, \citet{jung17} discussed that 
if the coverage of the anomaly is sparse, binary-source solutions can be 
confused with planetary solutions in the case of long-term planet-like 
perturbations.

In this paper, we demonstrate the importance of high-cadence and continuous 
observations, and the capability of second-generation microlensing experiments 
by presenting the analysis of the lensing event OGLE-2016-BLG-1003. 
The event was densely covered by the round-the-clock observations of 
three surveys including KMTNet, OGLE, and MOA. The light curve exhibits 
two nested pairs of caustic-crossing features, which is unusual for 
typical caustic-crossing perturbations.

\section{Observation}

\begin{figure}[th]
\epsscale{1.2}
\plotone{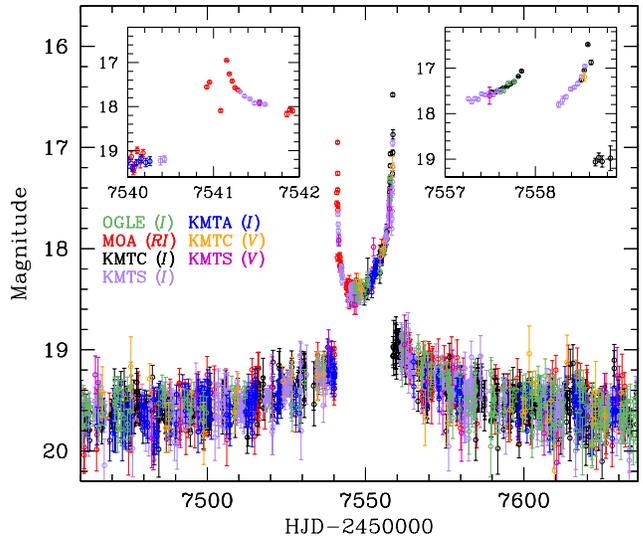}
\caption{\label{fig:one}
Light curve of OGLE-2016-BLG-1003. Two insets show the zoom of the 
perturbations centered at ${\rm HJD}' \sim 7541$ and $7558$, respectively. 
The notation in the parenthesis represents the observation passband.
}
\end{figure}

The equatorial coordinates of OGLE-2016-BLG-1003 are 
$(\alpha,\delta)_{\rm J2000} = (17^{\rm h}36^{\rm m}25^{\rm s}\hskip-2pt.23, 
-27^{\circ}11'19''\hskip-2pt.9)$, 
which correspond to Galactic coordinates $(l,b)=(0.41^\circ, 2.65^\circ)$. 
On 2016 June 18, the event was announced by the OGLE group through the Early 
Warning System \citep[EWS:][]{udalski15b}. The OGLE group uses the 1.3m Warsaw 
telescope at the Las Campanas Observatory in Chile. This field was independently 
monitored by the MOA survey with its 1.8m telescope at Mt.~John Observatory 
in New Zealand.

The lensed star was also in the fields of the KMTNet survey, which 
consists of three identical 1.6m telescopes equipped with a $4.0~\rm deg^{2}$ 
mosaic camera that are located at the Cerro Tololo Inter-American Observatory 
in Chile (KMTC), South African Astronomical Observatory in South Africa (KMTS), 
and Siding Spring Observatory in Australia (KMTA). With these wide-field telescopes 
that are globally distributed, the survey continuously observed the event 
with $\sim 1~\rm hr^{-1}$ cadence.

\begin{deluxetable}{lccc}
\tablecaption{Correction Parameters\label{table:one}}
\tablewidth{0pt}
\tablehead{
\multicolumn{1}{c}{Observatory}  &
\multicolumn{1}{c}{Number}       &
\multicolumn{1}{c}{$k$}          &
\multicolumn{1}{c}{$\sigma_{\rm min}$ (mag)}
}
\startdata
OGLE $(I)$  & 1716 &   1.587   & 0.005  \\
MOA $(RI)$  &  749 &   0.905   & 0.001  \\
KMTC $(I)$  &  791 &   0.922   & 0.001  \\
KMTS $(I)$  &  693 &   0.995   & 0.001  \\
KMTA $(I)$  &  290 &   1.077   & 0.003  \\
KMTC $(V)$  &   41 &   1.068   & 0.001  \\
KMTS $(V)$  &   20 &   1.015   & 0.001  
\enddata
\end{deluxetable}

\begin{figure}[th]
\epsscale{1.2}
\plotone{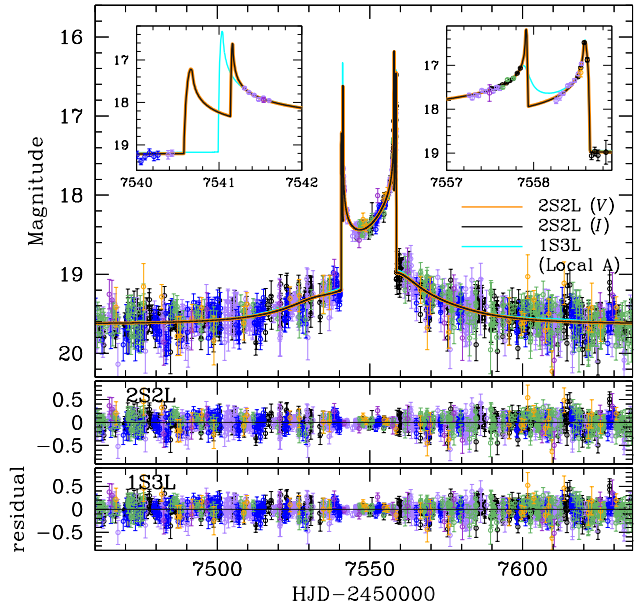}
\caption{\label{fig:two}
Light curves of 1S3L and 2S2L interpretations without the MOA data. 
The curve with cyan color is the best-fit 1S3L solution, while the two curves 
with orange and black colors are the best-fit 2S2L solutions for $V$ and $I$ band, 
respectively. The two insets show the zoom of the perturbations 
near ${\rm HJD}' \sim 7541$ and $7558$.
}
\end{figure}

Data reduction was performed by the individual survey groups using 
their customized pipelines. \footnote{For the source color measurement, 
the additional reduction of KMTC data was done using the DoPhot \citep{schechter93} pipeline.} 
All of these pipelines are developed based on 
the image subtraction technique \citep{alard98}. In order to use the data 
sets acquired from the different photometry codes, we readjust the error 
bars of each data set using the standard method in microlensing \citep{yee12}. 
Based on the error derived from the pipeline, $\sigma_{0}$, the remormalized 
error is calculated by 
\begin{equation}
\sigma^\prime = \sqrt{\sigma_{\rm min}^2 + (k\sigma_{0})^2}.
\label{eq1}
\end{equation}
Here $\sigma_{\rm min}$ is a factor needed to adjust the error to be consistent 
with the scatter of the data, and $k$ is a scaling factor needed to make 
$\chi^2/{\rm dof} \sim 1$. In Table~\ref{table:one}, we list the values of 
the error parameters for each data set with the total number of data 
and the observing passband.

\section{Analysis}

The OGLE-2016-BLG-1003 light curve (Figure~\ref{fig:one}) 
shows a large U-shaped brightness variation, 
suggesting that the event seems to be a typical binary-lens event 
characterized by a caustic entrance followed by a caustic exit. 
However, dense and continuous observations reveal that 
the light curve exhibits four distinctive spikes that 
occurred at ${\rm HJD}'(={\rm HJD} - 2,450,000) \sim 7540.9, 7541.2, 7557.9$, and $7558.6$. 
Such complex deviations are unusual for typical caustic-crossing perturbations.

In fact, the MOA data remained unreduced 
until we encountered a degeneracy between two interpretations 
as presented below. Although one of these was proved to be incorrect 
from the MOA data, examining the degeneracy is very important because 
(1) it raises the issue of the limitations of current analysis techniques 
for understanding lens systems beyond two masses (i.e., triple-lens system), 
and (2) it provides an opportunity for pondering the observational 
strategy of future microlensing experiments. Therefore, 
we first study the event without the MOA data.

\subsection{Without MOA data}

\begin{figure}[th]
\epsscale{1.2}
\plotone{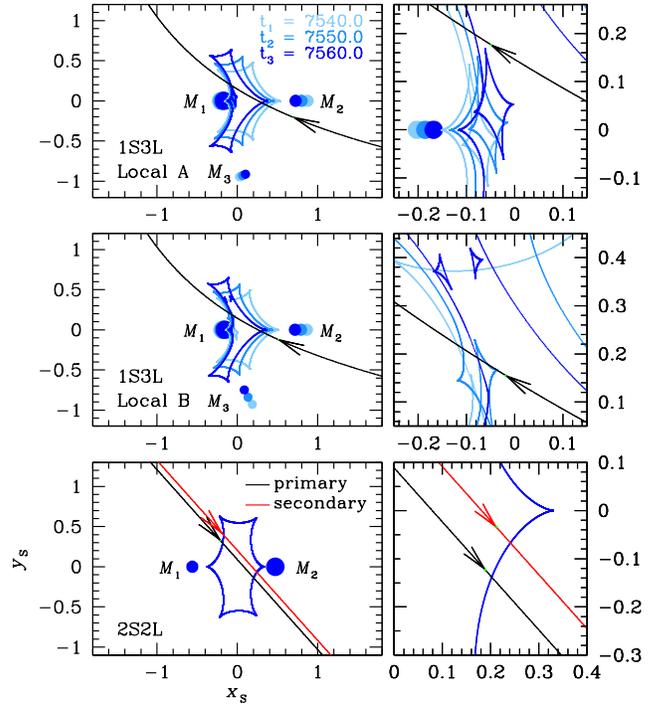}
\caption{\label{fig:three}
Geometries of 1S3L and 2S2L interpretations without the MOA data. 
In each panel, the closed curve represents the caustic, and the line with 
an arrow is the source trajectory. The locations of the lens components are 
represented by the blue filled circles. The right panels show the enlarged view 
of the left panel corresponding to the anomaly region near ${\rm HJD}' \sim 7558$. 
In the case of the 1S3L solutions, the positions of lens component and caustic 
change in time due to the orbital motion. We show three sets of 
caustics at ${\rm HJD}' \sim 7540$, $7550$, and $7560$. 
All lengths are normalized to the Einstein radius $\theta_{\rm E}$.
}
\end{figure}

To explain the light curve, we initially search for the solution based on a 
single-source binary-lens (1S2L) interpretation. Using the parametrization 
and the method of \citet{jung15}, we investigate the parameter space considering 
higher-order effects such as lens orbital motion \citep{dominik98} and 
microlens parallax \citep{gould92a,gould04}. However, 
because of the two distinct features at ${\rm HJD}' \sim 7557.9$ and $7558.6$, 
we could not find a satisfactory model describing the perturbations. 
We therefore take into account two additional interpretations 
that have a possibility to cause such complex deviations. 
The first interpretation is that the lens is composed of 
three masses (1S3L) and the second is that both the source 
and the lens contain two components (2S2L).

\subsubsection{Single-source Triple-lens Modeling}

The 1S3L modeling is conducted using a similar method to that of \citet{udalski15a} 
and \citet{bennett17}. First, we search for the 1S2L parameters that describe 
the dominant U-shaped anomaly by removing the data around ${\rm HJD}' \sim 7558.6$. 
We then recover the data and perform an additional grid search over $(s_{2}, q_{2}, \psi)$ 
space by fixing the $(s_{1}, q_{1})$ parameters, where $s_{1}$ 
(normalized to $\theta_{\rm E}$) is the projected binary separation 
and $q_{1} = M_{2}/M_{1}$. Here $s_{2}$ (normalized to $\theta_{\rm E}$) 
is the projected separation between the third mass $M_{3}$ and the barycenter of 
the binary lens $({\rm CM_{B}})$, $\psi$ is the angle between 
the ${\rm CM_{B}}-M_{3}$ axis and the binary-lens axis, and $q_{2} = M_{3}/M_{1}$. 
We note that the grid parameters $(s_{1}, q_{1}, s_{2}, q_{2}, \psi)$ are fixed during 
the computation, while other parameters are allowed to vary from their initial values. 
Finally, to refine each local minimum, we seed a Markov Chain Monte Carlo (MCMC) with 
its parameters, and allow all of these to vary.

\begin{deluxetable*}{l|rrrrr|rr}
\tablecaption{Best-fit Parameters\label{table:two}}
\tablewidth{0pt}
\tablehead
{
\multicolumn{1}{c|}{Parameters} &
\multicolumn{5}{c|}{Without MOA data} &
\multicolumn{2}{c}{With MOA data} \\
\multicolumn{1}{c|}{} &
\multicolumn{4}{c}{1S3L} &
\multicolumn{1}{c|}{2S2L} &
\multicolumn{2}{c}{2S2L} \\ 
\multicolumn{1}{c|}{} &
\multicolumn{2}{c}{Local A} &
\multicolumn{2}{c}{Local B} &
\multicolumn{1}{c|}{} &
\multicolumn{1}{c}{Without Prior} &
\multicolumn{1}{c}{With Prior} \\
\multicolumn{1}{c|}{} &
\multicolumn{1}{c}{Standard} &
\multicolumn{1}{c}{Orbit+Parallax} &
\multicolumn{1}{c}{Standard} &
\multicolumn{1}{c}{Orbit+Parallax} &
\multicolumn{1}{c|}{Standard} &
\multicolumn{1}{c}{Standard} &
\multicolumn{1}{c}{Standard} 
}

\startdata
$\chi^2$/dof                     &  3604.2/3541        &  3530.9/3535        &  3610.7/3541        &  3531.8/3535         &  3531.3/3539                  & 4288.1/4288        &                     \\
$t_{0,1}$ (HJD$'$)               &  7551.978$\pm$0.135 &  7552.916$\pm$0.137 &  7551.928$\pm$0.123 &  7552.931$\pm$0.153  &  7551.038$\pm$0.204           & 7549.825$\pm$0.182 & 7549.977$\pm$0.094  \\
$u_{0,1}$                        &  0.092$\pm$0.004    & -0.125$\pm$0.005    &  0.092$\pm$0.003    & -0.124$\pm$0.004     &  0.059$\pm$0.013              & -0.022$\pm$0.010   & -0.015$\pm$0.007    \\
$t_{0,2}$ (HJD$'$)               &  --                 &  --                 &  --                 &  --                  &  7552.517$\pm$0.134           & 7553.146$\pm$0.129 & 7553.256$\pm$0.107  \\
$u_{0,2}$                        &  --                 &  --                 &  --                 &  --                  &  0.135$\pm$0.006              & 0.135$\pm$0.004    & 0.134$\pm$0.003     \\
$t_{\rm E}$ (days)               &  39.832$\pm$1.415   &  31.971$\pm$1.056   &  40.216$\pm$1.345   &  33.334$\pm$1.030    &  28.931$\pm$0.665             & 29.860$\pm$0.632   & 30.203$\pm$0.649    \\
$s_1$                            &  0.954$\pm$0.014    &  1.066$\pm$0.013    &  0.948$\pm$0.013    &  1.042$\pm$0.014     &  1.033$\pm$0.011              & 1.014$\pm$0.010    & 1.006$\pm$0.011     \\
$q_1$                            &  0.248$\pm$0.013    &  0.232$\pm$0.011    &  0.249$\pm$0.012    &  0.217$\pm$0.014     &  1.188$\pm$0.039              & 1.407$\pm$0.013    & 1.428$\pm$0.014     \\
$\alpha$ (rad)                   &  2.554$\pm$0.015    & -2.652$\pm$0.016    &  2.551$\pm$0.013    & -2.654$\pm$0.018     &  0.842$\pm$0.015              & 0.875$\pm$0.014    & 0.893$\pm$0.012     \\
$s_2$                            &  0.941$\pm$0.011    &  0.944$\pm$0.021    &  0.786$\pm$0.011    &  0.937$\pm$0.021     &  --                           & --                 & --                  \\
$q_2$ $(10^{-3})$                &  1.054$\pm$0.162    &  1.256$\pm$0.158    &  1.082$\pm$0.192    &  1.176$\pm$0.251     &  --                           & --                 & --                  \\
$\psi$ (rad)                     &  1.594$\pm$0.028    & -1.538$\pm$0.043    &  1.615$\pm$0.032    & -1.377$\pm$0.051     &  --                           & --                 & --                  \\
$\rho_{*,1}$ $(10^{-3})$         &  0.603$\pm$0.058    &  0.685$\pm$0.062    &  0.553$\pm$0.056    &  0.570$\pm$0.060     &  $0.451_{-0.121}^{+1.422}$    & 1.003$\pm$0.351    & 1.321$\pm$0.048     \\
$\rho_{*,2}$ $(10^{-3})$         &  --                 &  --                 &  --                 &  --                  &  1.293$\pm$0.161              & 1.371$\pm$0.068    & 1.368$\pm$0.053     \\
$\pi_{{\rm E},N}$                &  --                 & -0.614$\pm$0.423    &  --                 & -0.585$\pm$0.367     &  --                           & --                 & --                  \\
$\pi_{{\rm E},E}$                &  --                 &  0.382$\pm$0.153    &  --                 &  0.603$\pm$0.110     &  --                           & --                 & --                  \\
$ds_{1}/dt$ $({\rm yr}^{-1})$    &  --                 & -3.342$\pm$0.122    &  --                 & -3.229$\pm$0.120     &  --                           & --                 & --                  \\
$d\alpha/dt$ $({\rm yr}^{-1})$   &  --                 &  1.273$\pm$0.179    &  --                 &  1.191$\pm$0.204     &  --                           & --                 & --                  \\
$ds_{2}/dt$ $({\rm yr}^{-1})$    &  --                 & -0.498$\pm$0.373    &  --                 & -3.526$\pm$0.405     &  --                           & --                 & --                  \\
$d\psi/dt$ $({\rm yr}^{-1})$     &  --                 &  1.500$\pm$0.443    &  --                 & -1.515$\pm$0.513     &  --                           & --                 & --                  \\
$q_{F,V}$                        &  --                 &  --                 &  --                 &  --                  &  1.202$\pm$0.201              & 1.047$\pm$0.234    & 1.070$\pm$0.036     \\
$q_{F,RI}$                       &  --                 &  --                 &  --                 &  --                  &  --                           & 0.981$\pm$0.032    & 1.033$\pm$0.031     \\
$q_{F,I}$                        &  --                 &  --                 &  --                 &  --                  &  1.182$\pm$0.054              & 1.037$\pm$0.037    & 1.068$\pm$0.036     
\enddata
\tablecomments{
${\rm HJD}'= {\rm HJD}-2450000$
}
\end{deluxetable*}

We find that two degenerate 1S3L models (marked as ``A'' and ``B'') can explain the 
light curve. However, there are some residuals in both solutions. 
This triggers the further investigation considering both the parallax and the lens 
orbital effect (``orbit+parallax''). Accounting for the parallax effect requires 
including two additional parameters 
${\pivec}_{\rm E} = (\pi_{\rm E, \it N}, \pi_{\rm E, \it E})$, 
which are the vector components of the microlens parallax \citep{gould04}. 
For the consideration of the orbital effects, we include four additional parameters 
$(ds_{1}/dt, d\alpha/dt, ds_{2}/dt, d\psi/dt)$, which describe, respectively, 
the change rate of $s_{1}$, $\alpha$, $s_{2}$, and $\psi$ to first-order approximation. 
When we test the model with these higher-order effects, we test $u_{0} > 0$ and 
$u_{0} < 0$ solutions to consider the ``ecliptic degeneracy'' \citep{jiang04,skowron11}. 
We note that because the third mass is much smaller than the other two masses, 
we adopt ${\rm CM_{B}}$ as a reference position of the lens system. 

In Table~\ref{table:two}, we present the two solutions. The best-fit model light curve 
(local A) is shown in Figure~\ref{fig:two}. The lensing geometries of the individual 
solutions are presented in the upper and middle panels of Figure~\ref{fig:three}. 
We find that the light curve is generated by a circumbinary planet orbiting a stellar 
binary. The perturbations near ${\rm HJD}' \sim 7541.2$ and $7558.6$ are produced by the 
source crossing over the resonant caustic formed by the binary lens, while the perturbation 
near ${\rm HJD}' \sim 7557.9$ is produced by the source passing close to the central 
caustic induced by the planet located near the Einstein ring. Although the two solutions 
cause ambiguity in the $s_{2}$ measurement, the two mass ratios $q_{1}$ and $q_{2}$ 
are almost same in both solutions.

\subsubsection{Binary-source Binary-lens Modeling}

The binary-source lensing magnification is described by the flux-weighted 
mean of two single-source magnifications, i.e., 
$A = (A_{1}F_{1}+A_{2}F_{2})/(F_{1}+F_{2})$, where $F_{i}$ and $A_{i}$ denote 
the flux and magnification of the individual sources \citep{griest92}. 
The single-source magnification is related to the lens mapping from the source 
plane to the image plane, resulting in the distortion of images. 
The lens-mapping equation is represented by 
\begin{equation}
\zeta = z - \sum_{i=1}^{N_{l}}{ \epsilon_i\over \bar{z}-\bar{z}_{m,i}},
\label{eq2}
\end{equation}
where $N_{l}$ is the number of lens components, $\epsilon_{i} = m_{i}/M_{\rm tot}$, 
and $M_{\rm tot}$ is the total mass of the lens system. Here $\zeta$, $z_{m,i}$, 
and $z$ are, respectively, the positions of the source, lens components, 
and images in complex coordinates, and the bar denotes the complex conjugate. 
Note that all angles are normalized to $\theta_{\rm E}$. 
The magnification $A_{j}$ of {\it j}th image is then determined from the amount 
of the distortion of the image given by the inverse of the determinant of 
the Jacobian, i.e.,
\begin{equation}
A_{j} = {1\over \left\vert{\rm det} J\right\vert}; \qquad
{\rm det} J= \left.{1 - {\partial\zeta \over \partial\bar{z}}
{\partial\bar{\zeta}\over \partial z}}\right\vert_{z=z_{j}},
\label{eq3}
\end{equation}
and the single-source magnification is thus the sum of all magnified 
images $A=\sum_j A_j$.

For the interpretation of the light curve, we define the principal 
2S2L lensing parameters with the approximation that the relative motion 
between the lens and each source star is rectilinear, and the trajectories 
of two sources are parallel each other. Based on the 1S2L parametrization, 
the description of a standard 2S2L light curve then requires 4 additional 
parameters related to the additional source companion: 
$(t_{0,2}, u_{0,2}, \rho_{*,2}, q_{F})$. Here the definitions of $(t_{0,2}, u_{0,2})$ 
are the same as those for a single-source single-lens event, $\rho_{*,2}$ is 
the source radius of the second source, and $q_{F} = F_{2}/F_{1}$ is 
the flux ratio between the two sources, which is needed to compute the individual 
source fluxes. We note that the flux ratio depends on the passband 
\citep{griest92,hwang13,jung17}, and thus one should allot separate 
$q_{F}$ parameters to each observed passband.

\begin{figure}[th]
\epsscale{1.2}
\plotone{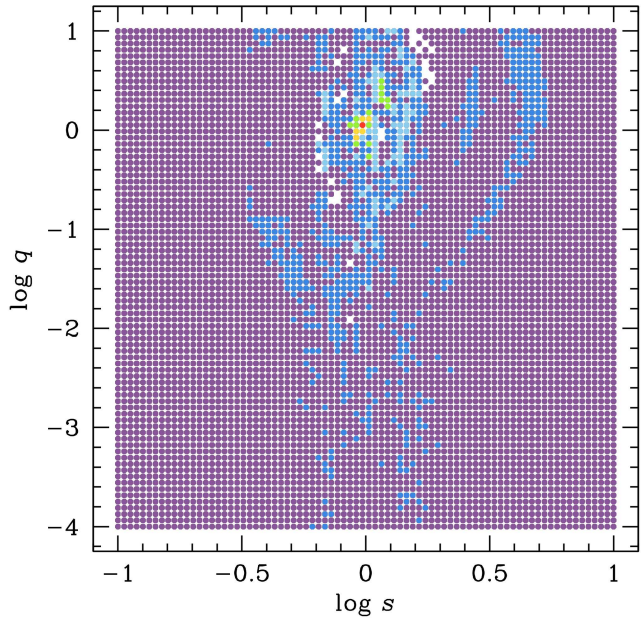}
\caption{\label{fig:four}
$\Delta\chi^{2}$ distribution in $(s,q)$ space obtained from the 2S2L grid search. 
Each color represents the point of the space within 
$1n\sigma$ (red), $2n\sigma$ (yellow), $3n\sigma$ (green), $4n\sigma$ 
(light blue), $5n\sigma$ (blue), and $6n\sigma$ (purple) level 
from the best-fit solution, where $n = 20$.           
}
\end{figure}

We search for the 2S2L solution using a method similar to \citet{jung15}. 
We first perform an initial grid search. We set $s$, $q$, $\alpha$, and $q_{F}$ 
as grid parameters since the lensing magnification varies sensitively to the 
change of these parameters, while the magnification varies smoothly to the 
changes of the other parameters. The grid parameters $(s, q, \alpha, q_{F})$ 
are divided by $(80, 80, 11, 11)$ grids and span the range of 
$-1.0 < {\rm log}~s < 1.0$, $-4.0 < {\rm log}~q < 1.0$, $0 < {\alpha} < 2\pi$, 
and $-5.0 < {\rm log}~q_{F} < 5.0$, respectively. At each grid point, we use the 
fixed value of $(s, q)$, while we allow $\alpha$ and $q_{F}$ to vary during the 
computation. Here we assume that the flux ratio is same for all passbands 
because the difference of magnification between different $q_{F}$ parameters 
is quite subtle compared to that of other parameters. Figure~\ref{fig:four} shows 
the $\Delta\chi^{2}$ distribution in the $(s,q)$ space obtained from this initial 
search. It clearly shows that there is only one local minimum. We then assign 
individual $q_{F}$ parameters to individual passbands, and find a best-fit 
solution from further refinement of the local solutions.

We find that the 2S2L solution also gives an excellent fit to the complex perturbations. 
In Table~\ref{table:two}, we list the best-fit parameters. The model light curve is presented 
in Figure~\ref{fig:two}. We show two model curves corresponding to each observed passband 
because the binary-source magnification is wavelength dependent. In Figure~\ref{fig:three}, 
we present the lensing geometry (lower panel) in which the two source trajectories are 
separately illustrated by a straight line with arrow head. We note that the source approaching 
closer to the barycenter of the lens is marked as the ``primary'' and the other source is 
marked as the ``secondary''. According to the 2S2L solution, the outer and inner pair of 
caustic-crossing features are produced by the secondary and the primary source 
passing over the resonant caustic, respectively.

\begin{figure}[th]
\epsscale{1.2}
\plotone{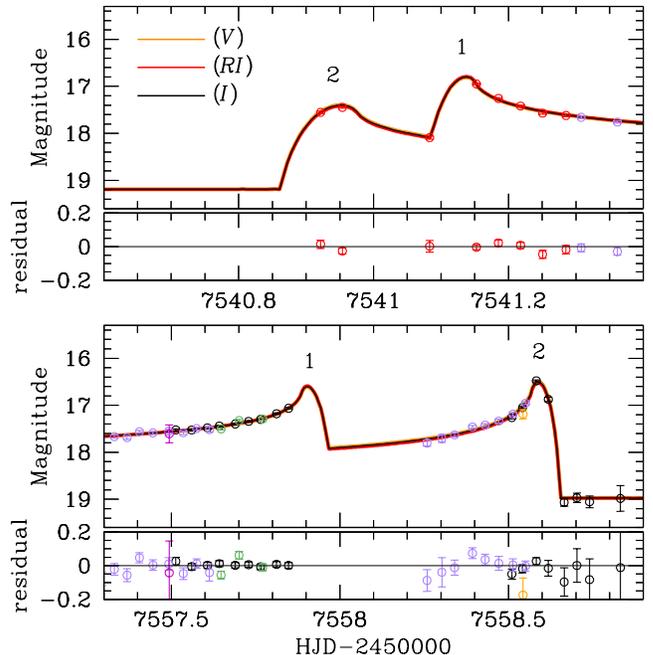}
\caption{\label{fig:five}
Enlarged view of the light curve around the caustic-crossing perturbations.
The three curves with orange, red, and black colors are the best-fit models 
for $V$, $RI$ and $I$ bands. They are nearly indistinguishable. 
The caustic entrance and the exit formed by the primary and the 
secondary source are marked by ``1'' and ``2'', respectively.
}
\end{figure}

From the comparison between the 1S3L and the 2S2L interpretations, we find that 
both interpretations almost equally well describe the observed light curve. 
The $\chi^{2}$ difference between two solutions is $\Delta\chi^{2} < 1$, indicating 
that they are extremely degenerate. The similarity of the fits notwithstanding, 
the two models predict very different brightening variation near 
${\rm HJD}' \sim 7541$: one caustic entrance for 1S3L model and two caustic 
entrances for 2S2L model. As a result, one could distinguish the two interpretations 
if there exist data points with a cadence high enough to identify such 
characteristic features.

\subsection{With MOA data}

The degeneracy seemed to remain unresolved. However, it was noticed that 
the event was also in one of the MOA observation fields, although it was not 
registered in the MOA event list. In addition, it was immediately recognized that 
the MOA data could play an important role to distinguish the degeneracy because 
the time zone of the MOA site exactly matched the unresolved part of the light curve.

We find that the MOA survey covers the unresolved region in which the data clearly 
show the feature of two caustic entrances. It definitively supports the 2S2L 
interpretation. With the MOA data, we further refine the 2S2L solution. 
The best-fit parameters are listed in Table~\ref{table:two}. 
In Figure~\ref{fig:five}, we present the light curve of the 2S2L solution in the 
caustic-crossing region. Since the overall light curve and the lensing geometry are 
almost same as those derived without the MOA data, we do not present these redundant figures. 
The estimated flux ratios $(q_{F,V}, q_{F,RI}, q_{F,I}) = (1.05, 0.98, 1.04)$ are 
very close to unity, implying that the brightness of the two sources is quite similar. 
This resemblance consequently indicates that the two sources would have similar $\rho_{*,i}$ values. 
However, unfortunately, we find that $\rho_{*,1}$ is not well constrained; it has a range of 
$0 < \rho_{*,1} < 0.0014$ due to the sparse coverage of the caustic crossing 
corresponding to the primary source (see Figure~\ref{fig:five}). 

\begin{figure}[th]
\epsscale{1.2}
\plotone{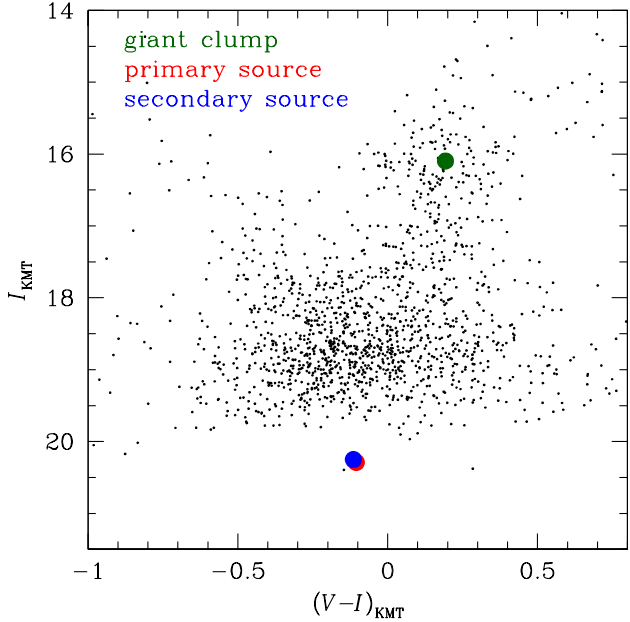}
\caption{\label{fig:six}
Positions of individual source stars in the instrumental color magnitude 
diagram of nearby stars. Also shown is the position of the giant clump centroid.  
}
\end{figure}

To check the result that comes from the poorly constrained $\rho_{*,1}$, 
we test an additional model including a physical constraint. 
One important characteristic of the 2S2L interpretation is that the measured 
$\theta_{\rm E,\it i}$ from the individual source radii $\rho_{*,i}$ must be 
consistent each other. Following this argument, we impose the gaussian constraint, 
\begin{equation}
f(\theta_{\rm E,1}) \sim e^{-(\theta_{\rm E,1}-\theta_{\rm E,2})^{2}/\sigma_{\rm E,2}^2},
\label{eq4}
\end{equation}
on the MCMC chains by adopting the error of $\theta_{\rm E}$ as 
$\sigma_{\rm E,2}$ (see next section). 
For the measurement of $\theta_{\rm E,\it i}$, 
we include the KMTC DoPhot $I$ and $V$ band data solely to obtain their 
flux fraction by fitting along with other data sets. 
As presented in the seventh and eighth columns of Table~\ref{table:two}, 
we find that the derived parameters are almost 
consistent and the prior does not affect the precision of the parameters significantly. 
In addition, the measured $\rho_{*,2}$ from both solutions are almost exactly same, indicating that 
the derived $\theta_{\rm E}$ is also consistent within the uncertainty level $\sigma_{\rm E}$. 
Therefore, we use the result to characterize the two source stars.

\subsection{Source Type}

We determine the source type by following the procedure of \citet{yoo04}. We first identify 
the individual source positions in the instrumental color-magnitude diagram (see Figure~\ref{fig:six}). 
We next measure the offsets $\Delta(V-I, I)$ between the giant clump centroid (GC) and the individual 
sources. From the de-reddened brightness $I_{0,\rm GC} = 14.43$ \citep{nataf13} and 
color $(V-I)_{0,\rm GC} = 1.06$ \citep{bensby11} of GC, we then estimate the de-reddened brightness 
$I_{0}$ and color $(V-I)_{0}$ of each source star. Here we assume that the GC and the source 
experience identical extinction. The derived $(V-I, I)_{0,i}$ of the two sources are 
$(V-I, I)_{0,1} = (0.76 \pm 0.13, 18.63 \pm 0.05)$ and 
$(V-I, I)_{0,2} = (0.75 \pm 0.12, 18.59 \pm 0.05)$, respectively. 
These indicate that the binary source is composed of two G-type main sequence stars.

Since we measure the finite size of the secondary source $\rho_{*,2}$, we also determine 
the angular Einstein radius $\theta_{\rm E}$. We first convert $(V-I)_{0,i}$ into 
$(V-K)_{0,i}$ using the $VIK$ color-color relation \citep{bessell88}, and then deduce 
the angular size of each source star $\theta_{*,i}$ from the relation between 
$(V-K)_{0}$ and $\theta_{*}$ \citep{kervella04}. The derived angular radii of the individual 
sources are $(\theta_{*,1}, \theta_{*,2}) = (0.629 \pm 0.045, 0.634 \pm 0.045)$ $\mu$as. 
The angular Einstein radius is thus 
\begin{equation}
\theta_{\rm E} = {\theta_{*,2} \over \rho_{*,2}} = 0.462 \pm 0.043~{\rm mas},
\label{eq5}
\end{equation}
and the proper motion of the binary source relative to the binary lens is    
\begin{equation}
\mu = {\theta_{\rm E} \over t_{\rm E}} = 5.654 \pm 0.521~{\rm mas~yr^{-1}}.
\label{eq6}
\end{equation}

\section{Discussion}

We have presented an analysis of the event OGLE-2016-BLG-1003 for which 
the light curve exhibits two pairs of caustic-crossing perturbations. 
From the analysis considering two different interpretations based on 
high-cadence and continuous data taken by three microlensing surveys, 
we found that the anomalies can be explained by the 2S2L model. 
Our 1S3L models were clearly disfavored by the MOA data. However, 
the analysis of the event and the process of reaching the conclusion that 
the 2S2L model is correct demonstrate the potential difficulties of 
interpreting such complex events.

First, our result does not fully guarantee that the 1S3L interpretation 
is not the solution, because we only searched a subset of 1S3L parameter 
space as described in Section 3.1.1. In fact, it is extremely difficult 
or almost impossible with our current level of understanding 
to conduct the full 1S3L analysis since current computing power 
is not even sufficient to obtain the initial solution by setting 
$(s_{1}, q_{1}, \alpha, s_{2}, q_{2}, \psi)$ as independent 
parameters. Therefore, in most cases, the 1S3L modeling has been done by following 
a hybrid approach that has been commonly used in this field. This approach is 
quite reasonable provided that the perturbation induced by the additional lens component 
does not affect the overall light curve (e.g., planetary perturbation). 
For example, \citet{udalski15a} found the 1S3L solution by first estimating 
the principal binary-lens parameters that explain the overall light curve, 
and then by adding the third mass to fit the planet-like perturbation 
similar to the approach used here. Experience from these events 
indicates that applying this approach to possible 1S3L 
candidates is extremely difficult if we could not first find the principal 
binary-lens parameters. Furthermore, even though the method can provide 
a plausible model for the light curve, there is no guarantee 
(i.e., $\Delta\chi^{2}$ surface) that the solution is the global minimum, 
and unfortunately this issue still remains an open question.

Our result also introduces degeneracy between 1S3L and 2S2L 
interpretations that can describe the unusual 
``nested caustics'' features (i.e., two caustic 
entrances or two caustic exits in a row). Although the event reported 
here favors the 2S2L interpretation, discriminating between 
them was only possible because we had high-cadence data and the two 
solutions predicted different lensing magnification curves, which will not always be the case. 
Nevertheless, it is possible to discriminate between 1S3L and 2S2L solutions 
due to the different origin of the lensing magnification. As already proven by 
analyzing observed events \citep{hwang13, jung17}, the binary-source magnification 
depends on the observing passband, and thus one can distinguish between two 
interpretations with multiband observations.

Until now, however, the major purpose of multiband observations from survey 
experiments was to determine the source color. Consequently, the observation 
cadence of the extra passband is much sparser than the primary passband. 
For example, the ratio between $V$ and $I$ band observations of 
KMTNet in 2016 was $1/11$ for CTIO site and $1/21$ for SAAO site. Although this 
cadence is high enough to obtain the source color, we found that it is insufficient 
to precisely estimate the flux ratio $q_{F,V}$ (see Table~\ref{table:two}), 
which yields the large uncertainty in the $V$ band magnification. As a result, 
it was difficult to identify the difference of the magnification between different passbands.

With second-generation experiments, we are now able to see many examples 
of complex perturbations such as reported here, and thus multiband observations 
become much more important to derive the correct solution. 
The microlensing event OGLE-2016-BLG-1003 proves the capability of 
current surveys to identify complex perturbations that are previously unknown. 
However, it also raises the issues of the limitation of current analysis and 
determining the appropriate multiband observing strategy of survey experiments.

\acknowledgments
This research has made use of the KMTNet system operated by the Korea 
Astronomy and Space Science Institute (KASI) and the data were obtained at 
three host sites of CTIO in Chile, SAAO in South Africa, and SSO in Australia. 
OGLE project has received funding from the National Science Centre, Poland, 
grant MAESTRO 2014/14/A/ST9/00121 to AU. The MOA project is supported by 
JSPS KAKENHI Grant Number JSPS24253004, JSPS26247023, JSPS23340064, 
JSPS15H00781, and JP16H06287. C. Han acknowledges support from Creative 
Research Initiative Program (2009-0081561) of National Research 
Foundation of Korea. A. Gould is supported from NSF grant AST-1516842 
and KASI grant 2016-1-832-01.


\begin{thebibliography}{99}

\bibitem[Alard \& Lupton(1998)]{alard98}
Alard, C., \& Lupton, Robert H. 1998, \apj, 503, 325

\bibitem[Alcock et al.(1993)]{alcock93}
Alcock, C., Akerlof, C. W., Allsman, R. A., et al. 1993, \nat, 365, 621

\bibitem[Aubourg et al.(1993)]{aubourg93}
Aubourg, E., Bareyre, P., Br\'{e}hin, S., et al. 1993, \nat, 365, 623

\bibitem[Bennett et al.(2017)]{bennett17}
Bennett, D. P., Rhie, S. H., Udalski, A., et al. 2017, \aj, 152, 125  

\bibitem[Bensby et al.(2011)]{bensby11}
Bensby, T., Ad\'{e}n, D., Mel\'{e}ndez, J., et al. 2011, \aap, 533, 134

\bibitem[Bessell \& Brett(1988)]{bessell88}
Bessell, M. S., \& Brett, J. M. 1988, \pasp, 100, 1134


\bibitem[Dominik(1998)]{dominik98}
Dominik, M. 1998, \aap, 329, 361

\bibitem[Dominik(1999)]{dominik99}
Dominik, M. 1999, \aap, 349, 108


\bibitem[Erdl \& Schneider(1993)]{erdl93}
Erdl, H., \& Schneider, P. 1993, \aap, 268, 453

\bibitem[Gaudi \& Han(2004)]{gaudi04}
Gaudi, B. S., \& Han, C. 2004, \apj, 611, 528

\bibitem[Gould(1992)]{gould92a}
Gould, A.\ 1992, \apj, 392, 442

\bibitem[Gould(1994)]{gould94}
Gould, A. 1994, \apjl, 421, L71

\bibitem[Gould(2004)]{gould04}
Gould, A. 2004, \apj, 606, 319

\bibitem[Gould \& Andronov(1999)]{gould99}
Gould, A., \& Andronov, N. 1999, \apj, 516, 236


\bibitem[Griest \& Hu(1992)]{griest92}
Griest, K., \& Hu, W. 1992, \apj, 397, 362



\bibitem[Hwang et al.(2013)]{hwang13}
Hwang, K.-H., Choi, J.-Y., Bond, I. A., et al. 2013, \apj, 778, 55

\bibitem[Jiang et al.(2004)]{jiang04}
Jiang, G., DePoy, D. L., Gal-Yam, A., et al. 2004, \apj, 617, 1307

\bibitem[Jung et al.(2015)]{jung15}
Jung, Y. K., Udalski, A., Sumi, T., et al. 2015, \apj, 798, 123

\bibitem[Jung et al.(2017)]{jung17}
Jung, Y. K., Udalski, A., Yee, J. C., et al. 2017, \aj, 153, 129

\bibitem[Kervella et al.(2004)]{kervella04}
Kervella P., Th\'{e}venin F., Di Folco E., S\'{e}gransan D., 2004, \aap, 426, 297

\bibitem[Kim et al.(2016)]{kim16}
Kim, S.-L., Lee, C.-U., Park, B.-G., et al. 2016, JKAS, 49, 37

\bibitem[Nataf et al.(2013)]{nataf13}
Nataf, D. M., Gould, A., Fouqu\'{e}, P., et al. 2013, \apj, 769, 88

\bibitem[Paczy\'{n}ski(1986)]{paczynski86}
Paczy\'{n}ski, B. 1986, \apj, 304, 1

\bibitem[Park et al.(2014)]{park14}
Park, H., Han, C., Gould, A., 2014, \apj, 787, 71


\bibitem[Schechter et al.(1993)]{schechter93}
Schechter, P. L., Mateo, M., \& Saha, A. 1993, \pasp, 105, 1342

\bibitem[Schneider \& Weiss(1986)]{schneider86}
Schneider, P., \& Weiss, A. 1986, \aap, 164, 237

\bibitem[Shin et al.(2016)]{shin16}
Shin, I.-G., Ryu, Y. H., Udalski, A., et al. 2016, JKAS, 49, 73

\bibitem[Skowron et al.(2011)]{skowron11}
Skowron, J., Udalski, A., Gould, A., et al. 2011, \apj, 738, 87

\bibitem[Sumi et al.(2011)]{sumi11}
Sumi, T., Kamiya, K., Bennett, D. P., et al. 2011, \nat, 473, 349

\bibitem[Sumi et al.(2016)]{sumi16}
Sumi, T., Udalski, A., Bennett, D. P., et al. 2016 \apj, 825, 112

\bibitem[Udalski et al.(1993)]{udalski93}
Udalski, A., Szymanski, M., Kaluzny, J., et al. 1993, Acta Astron., 43, 28

\bibitem[Udalski et al.(2015a)]{udalski15a} 
Udalski, A., Jung, Y. K., Han, C., et al. 2015a \apj, 812, 47 

\bibitem[Udalski et al.(2015b)]{udalski15b}
Udalski, A., Szyma\'{n}ski, M. K., \& Szyma\'{n}ski, G. 2015b, Acta Astron, 65, 1

\bibitem[Udalski et al.(2015c)]{udalski15c} 
Udalski, A., Yee, J. C., Gould, A., et al. 2015c, \apj, 799, 237

\bibitem[Yee et al.(2012)]{yee12}
Yee, J. C., Shvartzvald, Y., Gal-Yam, A., et al. 2012, \apj, 755, 102

\bibitem[Yoo et al.(2004)]{yoo04}
Yoo, J., DePoy, D. L., Gal-Yam, A., et al. 2004, \apj, 603, 139


\end{thebibliography}
\end{document}